\documentclass[byrevtex,nofootinbib,titlepage,nobibnotes,oneside,pra,a4paper]{revtex4}
\usepackage{amssymb}
\usepackage{amsmath}

\input{tcilatex}
\usepackage{graphicx}

\begin{document}

\title{Remark On Quantum Battle of The Sexes Game}
\author{{\small Jiangfeng Du}\thanks{%
Email: djf@ustc.edu.cn}{\small , Hui Li}\thanks{%
Email: lhuy@mail.ustc.edu.cn}{\small , Xiaodong Xu}\thanks{%
Email: xuxd@mail.ustc.edu.cn}{\small , Mingjun Shi, Xianyi Zhou, Rongdian Han%
}}
\address{{\small Laboratory of Quantum Communication and Quantum Computation}%
\\
{\small Department of Modern Physics, 230027}\\
{\small University of Science and Technology of China, Hefei, 230026,
P.R.China}}
\date{February, 19, 2001}

\begin{abstract}
Recently Quantum \textit{Battle of The Sexes Game }has been studied by Luca
Marinatto and Tullio Weber. Yet some important problems exist in their
scheme. Here we propose a new scheme to quantize \textit{Battle of The Sexes
Game}, and this scheme will truly remove the dilemma that exists in the
classical form of the game.
\end{abstract}

\maketitle

\section*{Introduction}

Quantum game and quantum strategies is a new born field. Although many novel
features have been discovered by researchers in the previous works\cite%
{1,2,3,4,5}, some problems also exist. In a recent article of Luca Marinatto
and Tullio Weber\cite{6}, they proposed a scheme to quantize the famous 
\textit{Battle of The Sexes Game}. In the usual exposition of this game,
Alice and Bob are trying to decide where to go on Saturday night. Alice
wants to attend the Opera, while Bob prefers to watch TV. And both players
would be happier to spend the evening together rather than apart. Both
players want to maximize their individual payoff. Table (\ref{Table 1})
indicates the payoffs of Alice and Bob. The first entry in the parenthesis
refers to Alice's payoff, and the second to Bob's. To satisfy the
preferences of the two players, the condition $\alpha >\beta >\gamma $ is
imposed.%
\begin{equation}
\begin{tabular}{|c|c|c|}
\hline
& Bob: $O$ & Bob: $T$ \\ \hline
Alice: $O$ & $\left( \alpha ,\beta \right) $ & $\left( \gamma ,\gamma
\right) $ \\ \hline
Alice: $T$ & $\left( \gamma ,\gamma \right) $ & $\left( \beta ,\alpha
\right) $ \\ \hline
\end{tabular}
\label{Table 1}
\end{equation}%
There are two Nash equilibrium $\left( O,O\right) $ and $\left( T,T\right) $
existing in the classical form of the game. Since there is no transfer of
information between two players, they face the dilemma in choosing between
two stable solutions. If the mismatch situation occurs, that is if one
player chooses strategy $O$ and the other chooses strategy $T$, then the
payoff to both players is $\gamma $ and it is the worst situation.

Luca Marinatto and Tullio Weber\cite{6} give an Hilbert structure to the
strategic spaces of the players, so allowing the existence of linear
combinations of classical strategies to be interpreted according to usual
formalism of orthodox quantum mechanics. They applied this formalism to the
study of the\textit{\ Battle of the Sexes Game}. The result is that if both
players are allowed to play entangled quantum strategies, the game has a
unique solution.

This article attracts the attention of S.C. Benjamin and he made a comment%
\cite{7} on it. In his comment two observations were made:`` Firstly, the
overall quantization scheme is fundamentally very similar to the scheme
proposed by Eisert et al.\cite{2}-the similarity is non-obvious because of
the very different use of the word `strategy' in the two approaches.
Secondly, we argue that the quantum \textit{Battle of the Sexes Game} does
not in fact have a unique solution, hence the players are still subject to a
dilemma.'' In the work of Luca Marinatto and Tullio Weber\cite{6}, the
initial strategy is set to $\frac{1}{\sqrt{2}}\left( \left| 00\right\rangle
+\left| 11\right\rangle \right) $, and `tactics' of the players are limited
to a probabilistic choice between applying the identity $I$ and $C=\sigma
_{x}$. So Benjamin pointed out ``this is a severe restriction on the full
range of quantum mechanically possible manipulation''. In the scheme of Luca
Marinatto and Tullio Weber\cite{6}, $p^{\ast }$ and $q^{\ast }$ are used to
represent the probability of Alice's and Bob's choosing $\sigma _{x}$
respectively. The authors get two profile tactics which satisfy the maximum
expected payoff of the players. One is $\left( p^{\ast }=0,q^{\ast
}=0\right) $ and the other is $\left( p^{\ast }=1,q^{\ast }=1\right) $.
Since the final entangled quantum `strategy' remains unchanged in either
case, so this is the unique solution in the game. However, Benjamin wrote
that ``it seems to us that a clear dilemma remains for the players''.
Because there are two pairs of tactics $\left( p^{\ast }=0,q^{\ast
}=0\right) $ and $\left( p^{\ast }=1,q^{\ast }=1\right) $, without knowing
the choice of the other player, the players face the dilemma to choose
between them, Benjamin continue, ``If the tactics are mismatched, i.e. if
either $\left( p^{\ast }=0,q^{\ast }=1\right) $ or $\left( p^{\ast
}=1,q^{\ast }=0\right) $ are adopted, then the worst situation will occur.
This is almost the same problem faced by the players in the traditional
game.''

In replying these remarks, Luca Marinatto \& Tullio Weber outlined some
topical points of their work which they hold to be crucial for a better
comprehension of their new approach to the quantum theory of games\cite{8}.
Regarding terminology, they think ``the choice of calling `strategies' the
quantum states instead of the `operators' used to manipulate them, is very
natural and quite consistent with the spirit of the classical game theory''.
As far as the choice of the tactics set is concerned, \ they wrote that
``obviously the class of allowed manipulations could be enlarged, but in our
paper we did not take care of this possibility since our minimal choice was
enough to reproduce intact the classical results when considering only
factorizable strategies and to obtain the disappearance of the dilemma when
resorting to entangled strategies''. As regards the possibility of the
occurrence of the mismatch situation, they consider it from the perspective
of practical operation. They think that ``since both the choices will
eventually lead to the same final strategy'', ``it is therefore apparent
that both players, knowing this fact, will decide doing nothing on their
strategy''. They have two reasons for their decision. One is that ``doing
nothing could be considered cheaper than doing something''. The other is
``Which player will in fact decide to refuse a certain gain, without
prospects of a better gain and running the risk of incurring a loss?''.

The reply of Luca Marinatto and Tullio Weber sounds reasonable\ from the
perspective of their paper. Yet we consider the terminology `\textit{%
strategy'} as the operation of a player on his state. The process of
obtaining initial state is not included in the strategic space. According to
the concept of the static game, the strategic moves of the players are local
operations, and each player has no information about the strategies of other
player. Since the process of obtaining initial state $\frac{1}{\sqrt{2}}%
\left( \left| OO\right\rangle +\left| TT\right\rangle \right) $ from $\left|
OO\right\rangle $ must be a global operation and every player clearly knows
it, so this process could not be called as \textit{strategy}. Furthermore we
think that in the scheme of Luca Marinatto \& Tullio Weber the probability $%
p^{\ast }$ and $q^{\ast },$ which are used to represent the probability of
Alice's and Bob's choosing $\sigma _{x}$ respectively, are classical
probabilities. Since $\sigma _{x}$ and $I$ are equivalent to classical pure
strategies ``flip'' and ``not-flip'' respectively, the operation of the
player could be considered as the classical mixture of classical strategies.

In this paper, we propose a new scheme in which there is no restriction on
the quantum mechanically possible manipulations of the players, and the
strategy space is the complete set of $SU\left( 2\right) $. Here the
terminology `strategy' is consistent with the one used by Eisert et al.\cite%
{2}. And the operation of a player is the probabilistic mixture of the pure
quantum strategies. We show that if both players resort to mixed quantum
strategy, although there are still many profiles of Nash equilibria, the
possible tactic mismatch will have no influence on the players. So the
dilemma which exists in the traditional \textit{Battle of the Sexes Game }is
removed.

\section{The Quantization of \textit{Battle of The Sexes Game}}

Our physical model for quantizing the game is similar to the delicate scheme
introduced by Eisert et al.\cite{2,9}. Figure 1 indicates the flow of
information in the \textit{Battle of the Sexes Game}.\FRAME{ftbpFU}{3.4921in%
}{1.9623in}{0pt}{\Qcb{The setup of Battle of Sexes Game.}}{\Qlb{Figure1}}{%
mixfig1.eps}{\raisebox{-1.9623in}{\includegraphics[height=1.9623in]{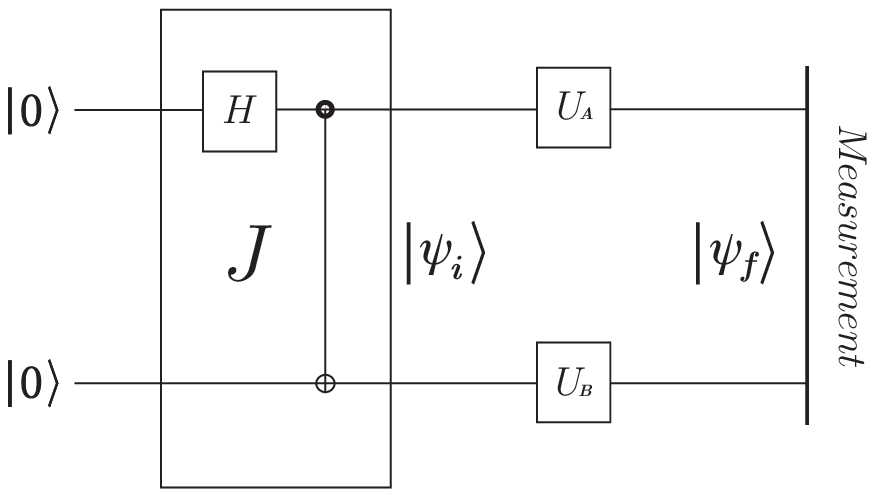}}}It is composed of a source of two
bits, one bit for each player. $\left| O\right\rangle =\left( 
\begin{array}{c}
1 \\ 
0%
\end{array}%
\right) $ represents the state of going to Opera and $\left| T\right\rangle
=\left( 
\begin{array}{c}
0 \\ 
1%
\end{array}%
\right) $ represents the state of watching TV. The state of the game is
described by a vector in the tensor product space which is spanned by the
basis $\left| OO\right\rangle $, $\left| OT\right\rangle $, $\left|
TO\right\rangle $ and $\left| TT\right\rangle $, where the first entry
refers to Alice' state and the second to Bob'. Gate $J$ is a unitary
operator and consist of a Hadamard gate and a C-NOT gate. So the prepared
state after gate $J$ is%
\begin{equation}
\left| \psi _{i}\right\rangle =J\left| OO\right\rangle =\frac{1}{\sqrt{2}}%
\left( \left| OO\right\rangle +\left| TT\right\rangle \right)  \label{eq 1}
\end{equation}%
So the density matrix of the initial state is 
\begin{equation}
\rho _{i}=\left| \psi _{i}\right\rangle \left\langle \psi _{i}\right|
\label{eq 3}
\end{equation}%
The strategies of Alice and Bob are denoted by unitary operator $U_{A}$ and $%
U_{B}$. Alice and Bob independently operate on her/his qubit with $U_{A}$
and $U_{B}$ respectively. $U_{A}$ and $U_{B}$ are chosen from the strategy
space $S$. Considering the possible strategies of the players, the strategy
space $S$ should be all of $SU(2)$, and the general expression of the
unitary operator is 
\begin{equation}
U\left( \theta ,\phi ,\psi \right) =\left( 
\begin{array}{cc}
e^{i\left( \phi +\psi \right) /2}\cos \frac{\theta }{2} & ie^{i\left( \phi
-\psi \right) /2}\sin \frac{\theta }{2} \\ 
ie^{-i\left( \phi -\psi \right) /2}\sin \frac{\theta }{2} & e^{-i\left( \phi
+\psi \right) /2}\cos \frac{\theta }{2}%
\end{array}%
\right)  \label{eq 2}
\end{equation}%
with $-\pi \leqslant \theta \leqslant \pi ,-\pi \leqslant \phi \leqslant \pi
,-\pi \leqslant \psi \leqslant \pi $. We set $f$ as the probability of the
operator $U$ which will be chosen by the players. Naturally $f$ is a
probability density function with respect to $U$. It is obvious that $%
f(U)\geqslant 0$ and $\dint\nolimits_{SU\left( 2\right) }f\left( U\right)
dU=1$. Here we use the invariance of the Haar measure, assumed to be
normalized so that the volume of $SU\left( 2\right) $ is $1$. The
probabilities of Alice choosing $U_{A}$ and Bob choosing $U_{B\text{ }}$are
denoted by $f_{A}(U_{A})$ and $f_{B}(U_{B})$ respectively. So after the
strategy moves of Alice and Bob, the final density matrix 
\begin{equation}
\rho _{f}=\dint_{A}\dint_{B}f_{A}f_{B}\left( U_{A}\otimes U_{B}\right) \rho
_{i}\left( U_{A}\otimes U_{B}\right) ^{+}dU_{A}dU_{B}  \label{eq 4}
\end{equation}%
The payoff operators for Alice and Bob are%
\begin{equation}
\left\{ 
\begin{array}{c}
\hat{\$}_{A}=\alpha \left| OO\right\rangle \left\langle OO\right| +\beta
\left| TT\right\rangle \left\langle TT\right| +\gamma \left( \left|
OT\right\rangle \left\langle OT\right| +\left| TO\right\rangle \left\langle
TO\right| \right) \\ 
\hat{\$}_{B}=\beta \left| OO\right\rangle \left\langle OO\right| +\alpha
\left| TT\right\rangle \left\langle TT\right| +\gamma \left( \left|
OT\right\rangle \left\langle OT\right| +\left| TO\right\rangle \left\langle
TO\right| \right)%
\end{array}%
\right.  \label{eq 5}
\end{equation}%
The expected payoffs of Alice and Bob are mean values of these operators and
hence are functional of $f_{A}$ and $f_{B}$.%
\begin{eqnarray}
\bar{\$}_{A}\left( f_{A},f_{B}\right) &=&Tr\left( \rho _{f}\hat{\$}%
_{A}\right)  \nonumber \\
&=&\dint_{A}\dint_{B}f_{A}f_{B}\left[ Tr\left( \left( U_{A}\otimes
U_{B}\right) \rho _{i}\left( U_{A}\otimes U_{B}\right) ^{+}\hat{\$}%
_{A}\right) \right] dU_{A}dU_{B}  \nonumber \\
&=&\dint_{A}\dint_{B}f_{A}f_{B}\$_{A}\left( U_{A},U_{B}\right) dU_{A}dU_{B}
\label{eq 7} \\
\bar{\$}_{B}\left( f_{A},f_{B}\right) &=&Tr\left( \rho _{f}\hat{\$}%
_{B}\right)  \nonumber \\
&=&\dint_{A}\dint_{B}f_{A}f_{B}\left[ Tr\left( \left( U_{A}\otimes
U_{B}\right) \rho _{i}\left( U_{A}\otimes U_{B}\right) ^{+}\hat{\$}%
_{B}\right) \right] dU_{A}dU_{B}  \nonumber \\
&=&\dint_{A}\dint_{B}f_{A}f_{B}\$_{B}\left( U_{A},U_{B}\right) dU_{A}dU_{B}
\label{eq 8}
\end{eqnarray}

Where $\$_{A}\left( U_{A},U_{B}\right) $ and $\$_{B}\left(
U_{A},U_{B}\right) $\ are the payoff functions of Alice and Bob when they
adopt the pure quantum strategies $U_{A}$ and $U_{B}$ respectively, the
expression of $\$_{A}$ and $\$_{B}$ are as follows:%
\begin{eqnarray}
\$_{A} &=&\alpha P_{OO}+\beta P_{TT}+\gamma \left( P_{OT}+P_{TO}\right)
=\left( \alpha -\gamma \right) P_{OO}+\left( \beta -\gamma \right)
P_{TT}+\gamma  \label{eq 9} \\
\$_{B} &=&\alpha P_{OO}+\beta P_{TT}+\gamma \left( P_{OT}+P_{TO}\right)
=\left( \beta -\gamma \right) P_{OO}+\left( \alpha -\gamma \right)
P_{TT}+\gamma  \label{eq 10}
\end{eqnarray}%
Since Alice and Bob adopt the pure quantum strategies, here $P_{\sigma \tau
}=\left| \left\langle \sigma \tau \right| \left( U_{A}\otimes U_{B}\right)
\left| \psi _{i}\right\rangle \right| ^{2}$ is the joint probability of the
final state collapses into the basis $\left| \sigma \tau \right\rangle $.%
\begin{equation}
P_{OO}=P_{TT}=\frac{1}{4}\left( 1+\cos \theta _{A}\cos \theta _{B}-\cos
\left( \psi _{A}+\psi _{B}\right) \sin \theta _{A}\sin \theta _{B}\right)
\label{eq 11}
\end{equation}%
From the expression of $P_{OO}$ and $P_{TT}$, we have 
\begin{align}
\$_{A}\left( U_{A},U_{B}\right) & =\$_{B}\left( U_{A},U_{B}\right)  \nonumber
\\
& =\frac{\alpha +\beta -2\gamma }{4}\left( 1+\cos \theta _{A}\cos \theta
_{B}-\cos \left( \psi _{A}+\psi _{B}\right) \sin \theta _{A}\sin \theta
_{B}\right) +\gamma  \nonumber \\
& =\frac{\alpha +\beta +2\gamma }{4}+\frac{\alpha +\beta -2\gamma }{4}\left(
\cos \theta _{A}\cos \theta _{B}-\cos \left( \psi _{A}+\psi _{B}\right) \sin
\theta _{A}\sin \theta _{B}\right)  \label{eq 12}
\end{align}

Substitute equation(\ref{eq 12}) into equations(\ref{eq 7},\ref{eq 8}), we
can obtain expressions of $\bar{\$}_{A}$ and $\bar{\$}_{B}$\ as funtionals
of $f_{A}^{\ast }$ and $f_{B}^{\ast }$.

Now we investigate Nash Equilibrium in this game. The definition of Nash
Equilibrium $\left\{ s_{A}^{\ast },s_{B}^{\ast }\right\} $ for pure strategy
in two-player games can be expressed as the following inequalities%
\begin{equation}
\left\{ 
\begin{array}{c}
\$_{A}\left( s_{A}^{\ast },s_{B}^{\ast }\right) \geqslant \$_{A}\left(
s_{A},s_{B}^{\ast }\right) \\ 
\$_{B}\left( s_{A}^{\ast },s_{B}^{\ast }\right) \geqslant \$_{B}\left(
s_{A}^{\ast },s_{B}\right)%
\end{array}%
\right. \forall s_{A},s_{B}\in SU\left( 2\right)  \label{eq 13}
\end{equation}%
When extended to the mixed strategies ($f_{A}$ and $f_{B}$), the Nash
Equilibrium profile $\left\{ f_{A}^{\ast },f_{B}^{\ast }\right\} $ can be
expressed as follows%
\begin{equation}
\left\{ 
\begin{array}{c}
\bar{\$}_{A}\left( f_{A}^{\ast },f_{B}^{\ast }\right) \geqslant \bar{\$}%
_{A}\left( f_{A},f_{B}^{\ast }\right) \\ 
\bar{\$}_{B}\left( f_{A}^{\ast },f_{B}^{\ast }\right) \geqslant \bar{\$}%
_{B}\left( f_{A}^{\ast },f_{B}\right)%
\end{array}%
\right. \forall f_{A}\geqslant 0,\forall f_{B}\geqslant
0,\int\nolimits_{A}f_{A}dU_{A}=\int\nolimits_{B}f_{B}dU_{B}=1.  \label{eq 14}
\end{equation}%
It is obvious that $f_{A}^{\ast }\geqslant 0,f_{B}^{\ast }\geqslant
0,\int\nolimits_{A}f_{A}^{\ast }dU_{A}=\int\nolimits_{B}f_{B}^{\ast
}dU_{B}=1 $. By using calculus of variations to the equation(\ref{eq 14}),
we obtain the following equations:%
\begin{equation}
\left\{ 
\begin{array}{c}
\left( \frac{\delta \bar{\$}_{A}\left( f_{A},f_{B}^{\ast }\right) }{\delta
f_{A}}\right) _{f_{A}=f_{A}^{\ast }}=\lambda _{A} \\ 
\left( \frac{\delta \bar{\$}_{B}\left( f_{A}^{\ast },f_{B}\right) }{\delta
f_{B}}\right) _{f_{B}=f_{B}^{\ast }}=\lambda _{B}%
\end{array}%
\right.  \label{eq 15}
\end{equation}%
where $\lambda _{A}$ and $\lambda _{B}$ are constant. From equation(\ref{eq
7},\ref{eq 8}), the left hand of the equation(\ref{eq 15}) can be rewritten
as follows%
\begin{equation}
\left\{ 
\begin{array}{c}
\left( \frac{\delta \bar{\$}_{A}\left( f_{A},f_{B}^{\ast }\right) }{\delta
f_{A}}\right) _{f_{A}=f_{A}^{\ast }}=\dint_{B}f_{B}^{\ast }\left(
U_{B}\right) \$_{A}\left( U_{A},U_{B}\right) dU_{B} \\ 
\left( \frac{\delta \bar{\$}_{B}\left( f_{A}^{\ast },f_{B}\right) }{\delta
f_{B}}\right) _{f_{B}=f_{B}^{\ast }}=\dint_{A}f_{A}^{\ast }\left(
U_{A}\right) \$_{B}\left( U_{A},U_{B}\right) dU_{A}%
\end{array}%
\right.  \label{eq 16}
\end{equation}%
From above calculation, we get the following equation.%
\begin{equation}
\left\{ 
\begin{array}{c}
\lambda _{A}=\frac{\delta \bar{\$}_{A}\left( f_{A},f_{B}^{\ast }\right) }{%
\delta f_{A}}=\dint_{B}f_{B}^{\ast }\left( U_{B}\right) \$_{A}\left(
U_{A},U_{B}\right) dU_{B} \\ 
\lambda _{B}=\frac{\delta \bar{\$}_{B}\left( f_{A}^{\ast },f_{B}\right) }{%
\delta f_{B}}=\dint_{A}f_{A}^{\ast }\left( U_{A}\right) \$_{B}\left(
U_{A},U_{B}\right) dU_{A}%
\end{array}%
\right.  \label{eq 17}
\end{equation}%
Here, the left hand of equation $\lambda _{A}$ (or $\lambda _{B}$) is a
constant independent of $U_{A}$ and $U_{B}$, but the right hand of equation(%
\ref{eq 17}) generally depends on $U_{A}$ (or $U_{B}$). The equation has o
solution without the guarantee of $f_{B}^{\ast }\left( U_{B}\right) $ (or $%
f_{A}^{\ast }\left( U_{A}\right) $) that there is no $U_{A}(U_{B})$ in the
result of the integral of $\$_{A}\left( U_{A},U_{B}\right) $ (or $%
\$_{B}\left( U_{A},U_{B}\right) $). On the other hand, any $f_{A}^{\ast
}\left( U_{A}\right) $ and $f_{B}^{\ast }\left( U_{B}\right) $ which meet
the guarantee of $f_{B}^{\ast }\left( U_{B}\right) $ (or $f_{A}^{\ast
}\left( U_{A}\right) $) will be solution of equation(\ref{eq 17}), and the
profile $\{f_{A}^{\ast },f_{B}^{\ast }\}$ will be one of the Nash Equilibria
of the game. When Alice's mixed strategy is $f_{A}^{\ast }$ and Bob's is $%
f_{B}^{\ast }$, neither of them can increase her/his payoff by unilaterally
changing her/his own strategy.

The solutions of equation (\ref{eq 17}) are obviously infinite, that means
there is no unique stable strategies. However, the possible tactic mismatch
will have no influence on the players because for any $f_{A}^{\ast }$ which
satisfies equation(\ref{eq 17}) will yield 
\begin{eqnarray}
\bar{\$}_{A}\left( f_{A}^{\ast },f_{B}^{\ast }\right)
&=&\dint_{A}\dint_{B}f_{A}^{\ast }f_{B}^{\ast }\$_{A}\left(
U_{A},U_{B}\right) dU_{A}dU_{B}  \nonumber \\
&=&\dint_{A}f_{A}^{\ast }\left[ \dint_{B}f_{B}^{\ast }\$_{A}\left(
U_{A},U_{B}\right) dU_{B}\right] dU_{A}  \nonumber \\
&=&\lambda _{A}\dint_{A}f_{A}^{\ast }dU_{A}  \label{eq 19} \\
&=&\lambda _{A}
\end{eqnarray}%
So the payoff for Alice ($\bar{\$}_{A}\left( f_{A}^{\ast },f_{B}^{\ast
}\right) $) is not related to her own strategy ($f_{A}^{\ast }$) but her
opponent's strategy($f_{B}^{\ast }$). Similarly the payoff for Bob ($\bar{\$}%
_{B}\left( f_{A}^{\ast },f_{B}^{\ast }\right) $) is $\lambda _{B}$, which is
not related to his own strategy ($f_{B}^{\ast }$). It is apparent that for
any combination of $f_{A}^{\ast }$ and $f_{B}^{\ast }$ which belongs to the
stable strategies (solutions of equation(\ref{eq 17})), the payoffs for
alice is $\lambda _{A}$ and for Bob is $\lambda _{B}$. Further more, through
the following caculation we find that the payoffs of Alice and Bob is the
same ($\lambda _{A}=\lambda _{B}$). So the players will not worry about the
occurrence of tactic mismatch. Then the dilemma in the classical \textit{%
Battle of \ The Sexes Game} is removed.

From the expression of $\theta ,\phi ,\psi $ in equation(\ref{eq 2}), the
normalized invariance of the Haar measure is 
\begin{equation}
dU\left( \theta ,\phi ,\psi \right) =\frac{1}{16\pi ^{2}}\left| \sin \theta
\right| d\theta d\phi d\psi  \label{eq 18}
\end{equation}%
So the integration on the right hand of the equation(\ref{eq 17}) can be
rewritten as integration with respect to $\theta ,\phi ,\psi $. For any $%
f_{A}^{\ast }$ which satisfies the equation(\ref{eq 17}), we get $%
\dint_{A}f_{A}^{\ast }\left( U_{A}\right) \$_{B}\left( U_{A},U_{B}\right)
dU_{A}$ being independent of $U_{B}$, hence $\dint_{A}f_{A}^{\ast }\left(
U_{A}\right) P_{\sigma \tau }\left( U_{A},U_{B}\right) dU_{A}$ is
independent of $U_{B}$. From the expression of $P_{\sigma \tau }$(see
equation(\ref{eq 11})),after integrating with respect to $\theta _{A},\phi
_{A},\psi _{A}$, the values of the integration of the terms that contain $%
\theta _{B},\phi _{B},\psi _{B}$ is zero. So we get 
\begin{eqnarray}
&&\dint_{A}f_{A}^{\ast }\left( U_{A}\right) P_{\sigma \tau }\left(
U_{A},U_{B}\right) dU_{A}  \nonumber \\
&=&\dint_{A}f_{A}^{\ast }\left( U_{A}\right) P_{OO}\left( U_{A},U_{B}\right)
dU_{A}  \nonumber \\
&=&\dint_{A}f_{A}^{\ast }\left( U_{A}\right) P_{TT}\left( U_{A},U_{B}\right)
dU_{A}  \nonumber \\
&=&\dint_{A}f_{A}^{\ast }\left( U_{A}\right) \tfrac{1}{4}dU_{A}=\frac{1}{4}
\label{eq 21}
\end{eqnarray}%
Repeating the same procedure, we obtain that $\dint_{B}f_{B}^{\ast }\left(
U_{B}\right) P_{\sigma \tau }\left( U_{A},U_{B}\right) dU_{B}=\dfrac{1}{4}$.
According to the payoff functions of $\$_{A}\left( U_{A},U_{B}\right) $ and $%
\$_{B}\left( U_{A},U_{B}\right) $ (see equation(\ref{eq 12})), finally the
payoff to both players is $\bar{\$}_{A}=\bar{\$}_{B}=\frac{\alpha +\beta
+2\gamma }{4}$. We list four of the simplest solutions(Nash Equilibria) in
the following table: 
\[
\begin{tabular}{|c|c|c|c|}
\hline
$\left( 1\right) $ & $f_{A}^{\ast }=1$ & $f_{B}^{\ast }=1$ & $\bar{\$}_{A}=%
\bar{\$}_{B}=\frac{\alpha +\beta +2\gamma }{4}$ \\ \hline
$\left( 2\right) $ & $f_{A}^{\ast }=1$ & $f_{B}^{\ast }=\frac{2}{\pi \left|
\sin \theta _{B}\right| }$ & $\bar{\$}_{A}=\bar{\$}_{B}=\frac{\alpha +\beta
+2\gamma }{4}$ \\ \hline
$\left( 3\right) $ & $f_{A}^{\ast }=\frac{2}{\pi \left| \sin \theta
_{A}\right| }$ & $f_{B}^{\ast }=1$ & $\bar{\$}_{A}=\bar{\$}_{B}=\frac{\alpha
+\beta +2\gamma }{4}$ \\ \hline
$\left( 4\right) $ & $f_{A}^{\ast }=\frac{2}{\pi \left| \sin \theta
_{A}\right| }$ & $f_{B}^{\ast }=\frac{2}{\pi \left| \sin \theta _{B}\right| }
$ & $\bar{\$}_{A}=\bar{\$}_{B}=\frac{\alpha +\beta +2\gamma }{4}$ \\ \hline
\end{tabular}%
\]%
Any strategy in the table is a stable solution.

In the classical \textit{Battle of The Sexes Game}, there are two Nash
Equilibria: $\left( O,O\right) $ and $\left( T,T\right) $. If the tactic
mismatch occurs, the strategic profile becomes $\left( O,T\right) $ or $%
\left( T,O\right) $ and the payoffs turn out to be the worst-case payoff $%
\bar{\$}_{A}=\bar{\$}_{B}=\gamma $. In the quantum version of this game,
there are more than one Nash Equilibrium appearing when players adopted the
deterministic strategies (see in Ref\cite{6}), and any tactic mismatch can
lead to the worst-case payoffs. While when we propose that if the players
adopt quantum mixed strategies, it can be guaranteed that the possible
appearance of the tactic-mismatch will have no effect on the payoffs of the
players, and the payoffs will always remain as $\frac{\alpha +\beta +2\gamma 
}{4}$.

\section{Conclusion}

In this paper, the problems existing in the previous quantum \textit{Battle
of The Sexes Game} are studied in detail. We proposed a new scheme, applying
mixed quantum strategies, to quantize this game. In our scheme, the players
can choose their strategies in the all of $SU\left( 2\right) $. It means
that there is no restriction on the quantum mechanically possible
manipulations of the players. We show that if the players resort to the
mixed quantum strategy, in the game the equilibria that they reach will be
much more efficient than that in its classical version. Furthermore, the
tactic-mismatch which is the difficulty faced by the players in the
traditional game\cite{7} has no effect on the payoffs of the players. Thus
the dilemma which exists in the classical \textit{Battle of the Sexes Game }%
is truly removed. The scheme of mixed quantum strategy proves successful. We
also hope that the scheme of mixed quantum strategy can be useful for the
investigation of other quantum games.

{\Large Acknowledgement}

This project was supported by the National Nature Science Foundation of
China (No. 10075041 and No. 10075044) and the Science Foundation for Young
Scientists of USTC.

\end{document}